\documentclass[%
reprint,
aps,
notitlepage,
twocolumn,
amsmath,
]{revtex4-2}
\usepackage{amsmath}
\usepackage{bm}

\usepackage{amsfonts}
\usepackage{xcolor} 
\usepackage{hyperref} 
\hypersetup{
	colorlinks   = true, 
	urlcolor     = red, 
	linkcolor    = blue, 
	citecolor   = blue 
}
\usepackage{graphicx}
\usepackage{cleveref}
\usepackage{physics}
\usepackage{geometry}
\newcommand{\AF}[1]{\textcolor{blue}{#1}}

\begin{document}
	\author{Wagner Tavares Buono}
		\email{wagner.tavaresbuono@wits.ac.za}
	\author{Jaqcuqueline Tau}
	\author{Isaac Nape}
	\author{Andrew Forbes}

	\affiliation{%
		Institute of Physics, University of the Witwatersrand
	}%
		\title{Eigenmodes of aberrated systems: the tilted lens}
\begin{abstract}

When light is passed through aberrated optical systems, the resulting degradation in amplitude and phase has deleterious effects, for example, on resolution in imaging, spot sizes in focussing, and the beam quality factor of the output beam. Traditionally this is either pre- or post-corrected by adaptive optics or phase conjugation.  Here we consider the medium as a complex channel and search for the eigenmodes of the channel, the modes that propagate through this system without alteration.  We employ a quantum-inspired approach and apply it to the tilted lens as our example channel, a highly astigmatic system that is routined used as a desired distortion inducer to measure orbital angular momentum.  We find the eigenmodes analytically, show their robustness in a practical experiment, and outline how this approach may be extended to arbitrary astigmatic systems. 

\end{abstract}
\maketitle


\section{Introduction}
Transverse or spatial modes that are solutions of the paraxial Helmholtz wave equation in vacuum are invariant through free-space propagation, where the term free-space refers to propagation in vacuum or a uniform non-changing index of refraction. Their orthogonality properties make them possible to be represented as vectors and matrices \cite{Allen1992,Allen1999} similar to the Jones formalism for polarization except for having increased dimensions. In cylindrical and Cartesian coordinates we have the famous Laguerre-Gaussian (LG) and Hermite-Gaussian (HG) modes, respectively. The modal description of light can be a powerful tool, including forming a measurement basis \cite{pinnell2020modal}, for control of beam sizes \cite{Carter1980,Phillips1983} to optical propagation description \cite{Sroor2021} and self-imaging selection rules \cite{Silva2020}. Based on their constituting polynomials, it is possible to decompose them into one another and thus describe mode conversion in terms of constituting polynomials \cite{Beijersbergen1993} but also matrices \cite{ONeil2000}.  These exotic forms of structured light \cite{forbes2021structured} have found many applications, from quantum \cite{forbes2019quantum} to classical \cite{rosales2018review}.  The modal description of light has in turn led to the search for the eigenmodes of the systems that the light propagates through, finding the modes that are invariant through a system.  This has successfully been achieved for optical eigenmode imaging \cite{Mazilu2011,Luca2011,Mazilu2011a}, super resolution \cite{Piche2012} and in microcavities \cite{Kaliteevski2000}, but often the orthogonality has been overlooked \cite{Kosmeier2014}. 

\vspace{0.5cm}

In this manuscript we use a quantum-inspired operator formalism to generate a matrix that describes our channel up until an arbitrary order. We use the tilted lens as our example of an aberrated system because of its extensive use in the sorting, detection and analysis of structure light carrying orbital angular momentum \cite{Allen1992}, making it a highly topical example.  From this matrix and operator it is immediate to see that HG modes are the eigenmodes. We find a general framework for understanding the full eigenmode set, including superpositions grouped by eigenvalue class, and confirm the findings experimentally by creating these complex patterns of light and verifying their resilient through the channel. Although we treat the highly astigmatic tilted lens as our topical example, the framework outlined here will be of value for further studies of arbitrary aberrated optical channels.

\section{Theory}

\subsection{The tilted lens}
The tilted lens emerged as a useful OAM detector, converting OAM modes to HG modes in a single plane based on its rotation (tilting) \cite{Vaity2013}.  It was a simplified version of an early approach based on cylindrical lenses \cite{Beijersbergen1993}, now known as astigmatic mode converters, and can be described by connecting wave and ray optics \cite{Collins1970lens,Shakir2015efficient}. The evolution of LG beams under astigmatism has been investigated in \cite{Wada2005} and \cite{Dennis2006} with the latter observing vortex splitting and elliptical intensity patterns similar to Ince-Gaussian modes \cite{Bandres2004}.  Mode transformation between LG and HG modes has been analyzed regarding OAM conservation \cite{Allen1992}, geometrical phase \cite{Galvez2003,Galvez2005}, vector beam transformations \cite{MendozaHernandez2019} and as an OAM and radial mode detector \cite{liu2021probing}. It has also been performed holographically \cite{Syouji2010} and used to identify superpositions of transverse modes in combination with machine learning \cite{Silva2021}.  Henceforth we will refer to the tilted lens as an astigmatic mode converter.

\vspace{0.5cm}

Because HG modes feature strongly in the application of such astigmatic mode converters \cite{ONeil2000,Vaity2013,Silva2021}, we use this as our basis for the modes and the operator.  Let us consider the vector $ \ket{\Psi} $ composed of every possible bi-dimensional Hermite-Gaussian transverse mode from the zero-th order up until the $ N_{max} $-th order as in
\begin{equation}
	\ket{\Psi}=\sum_{o=0}^{N_{max}} \sum_{n=0}^{o} c^n_{o-n} \ket{HG^n_{o-n}},
\end{equation}
with the spatial domain projection defined as
\begin{equation}\label{key}
	\braket{x,y}{HG^n_{m}}=H_m(\tilde{x})H_n(\tilde{y})e^{ikz}e^{i\phi_{m,n}(z)}e^{\frac{ik(x^2+y^2)}{R(z)}}G(\tilde{x},\tilde{y}),
\end{equation}
where $ \tilde{x}(\tilde{y})=x/w(z) (y/w(z)) $, the Gouy phase is $ \phi_{m,n}=(m+n+1)\arctan (z/z_R) $, $ w_0 $ is the beam waist, $ z_R $ is the Rayleigh length and $ G(\tilde{x},\tilde{y}) $ is the Gaussian envelope.

Similar to the Jones formalism, in the modal space, we can describe the action of one or more optical elements as an operator $ \mathcal{M} $ which operates in any state vector $ \ket{\Psi} $ as in
\begin{equation}
	\ket{\Psi '}=\mathcal{M}\ket{\Psi}
\end{equation}
and if the vector $ \ket{\Psi '}=\lambda\ket{\Psi} $ where $ \lambda $ is a complex scalar, then $ \ket{\Psi '} $ is an eigenvector of this operator. 

\vspace{0.5cm}
Remaining in the HG basis for describing the operator, we may write it compactly as
\begin{equation}\label{eq:operator}
	\mathcal{M}_t=\sum_{o}^{N_{max}}\sum_{n=0}^{o}e^{\frac{i(o-2n)\pi}{4}}\ket{HG^n_{o-n}}\bra{HG^n_{o-n}} \ .
\end{equation}
This operator defines \AF{the propagation of an optical mode to the tilted lens by a distance $f$ (where $f$ is the focal length of the untilted lens)}, the action of the lens itself, which is rotated around the $ z $ axis by a small angle, and the subsequent propagation after the tilted lens by a distance $ d $, to what we will call the conversion plane. 

\vspace{0.5cm}

From the definition of $ \mathcal{M}_t $ it is immediate to see that HG modes in the $x/y$ coordinates are all eigenvectors of this operator, independent of the order. However, since there are an infinite number of mode with discrete indices and their eigenvalues are given by a periodic function that increases with mode order, only a limited number of eigenvalues will be possible in this system. Therefore, regarding this transformation, an infinite number of HG modes can be grouped by a finite number of eigenvalues. This brings us to the often overlooked property of eigenvectors: \textit{any superposition of eigenvectors of the same eigenvalue is also an eigenvector}.

\begin{figure}[h!]
	\centering
	\includegraphics[width=\linewidth]{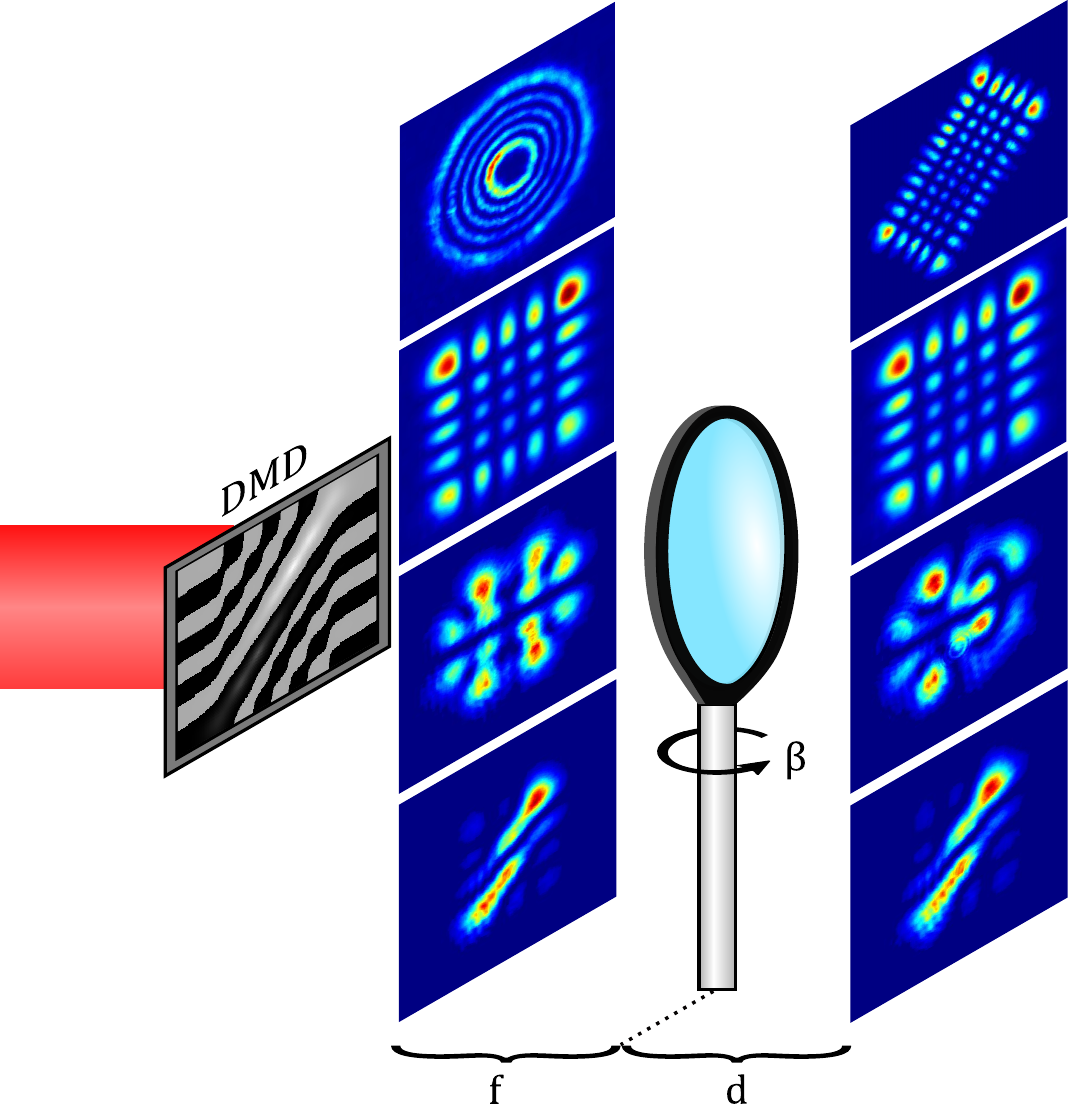}\caption{Laguerre-Gaussian modes are converted in HG modes when passed through a tilted lens (first row) after a propagation distance $ d $. HG modes in $ x/y $ directions are insensible to this transformation (second row) but an arbitrary combination of them is not (third row). A superposition of modes following the equivalence relation defined in \cref{eq:equivalence} is resilient to this aberration (forth row).\label{fig:concept}}
\end{figure}

We can grasp the concept in \cref{fig:concept}. The astigmatic mode converter can turn an LG beam into a HG mode in a propagation plane. This transformation does not change pure HG modes, but it can change an arbitrary superposition of them. However, by choosing a superposition of modes grouped by their eigenvalues, we generate states that are invariant through this astigmatic system.

\subsection{Generalization of Eigenvalues}

Since HG modes make a countable basis and given the form of the operator in \cref{eq:operator}, we can define $ j\in \{0,7\}$. We can see that there will be only 8 possible eigenvalues for this operator, where $ \lambda_j=e^{\frac{ij\pi}{4}} $ is the $ j-th $ eigenvalue. 

\begin{figure}[h!]
	\centering
	\includegraphics[width=\linewidth]{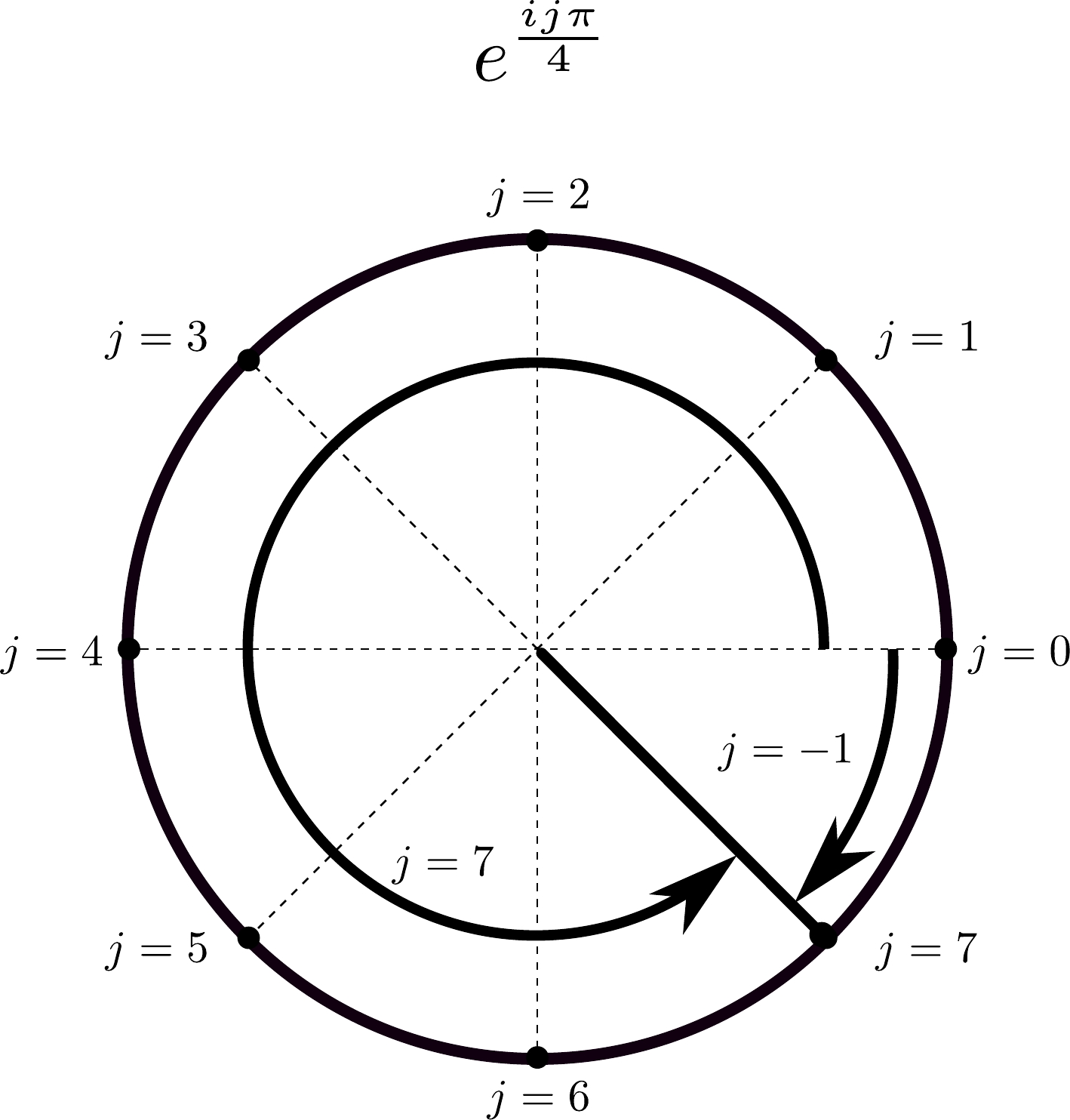}\caption{Eigenvalues mapping. The negative values can be mapped into positives using the periodicity of the exponential function.\label{fig:angles}}
\end{figure}

\vspace{0.5cm}

Notably, $ j $ can assume negative values, but given the periodicity of the exponential function, we can use modular arithmetic to show that any negative values can be mapped to positive ones according to $ j_{pos}=8\alpha+j_{neg} $, where $ \alpha $ is a positive integer. As an example, the value $ j=-1 $ would give the same eigenvalue as $ j=7 $, depicted in \cref{fig:angles}, or a value of $ j=-15 $ can be mapped to $ j=1 $. This can be summarized by the equivalence class

\begin{equation}\label{eq:equivalence}
	[j]= \{ \forall \ m,n \in \mathbb{W} \ | \ (m-n) \equiv j \ (\mathrm{mod} \ 8) \}
\end{equation}

where the expression $ a\equiv b (\mathrm{mod} \ 8) $ denotes a congruence module $ 8 $ relation \cite{epp2010discrete} and $ \mathbb{W} $ is the set of whole numbers (non-negative integers including zero \cite{koshy2002elementary}). An important property of this equivalence class is that it is not symmetric in any set of indices except for $ [0] $ and $ [4] $. A curious aspect is that the change from a positive $ [j] $ to its negative counterpart means a $ 90^o $ rotation around the $ z$ axis in the intensity patterns, which directly links, for example, the classes $ [3] $ and $ [5] $.

\begin{figure}[h!]
	\centering
	\includegraphics[width=\linewidth]{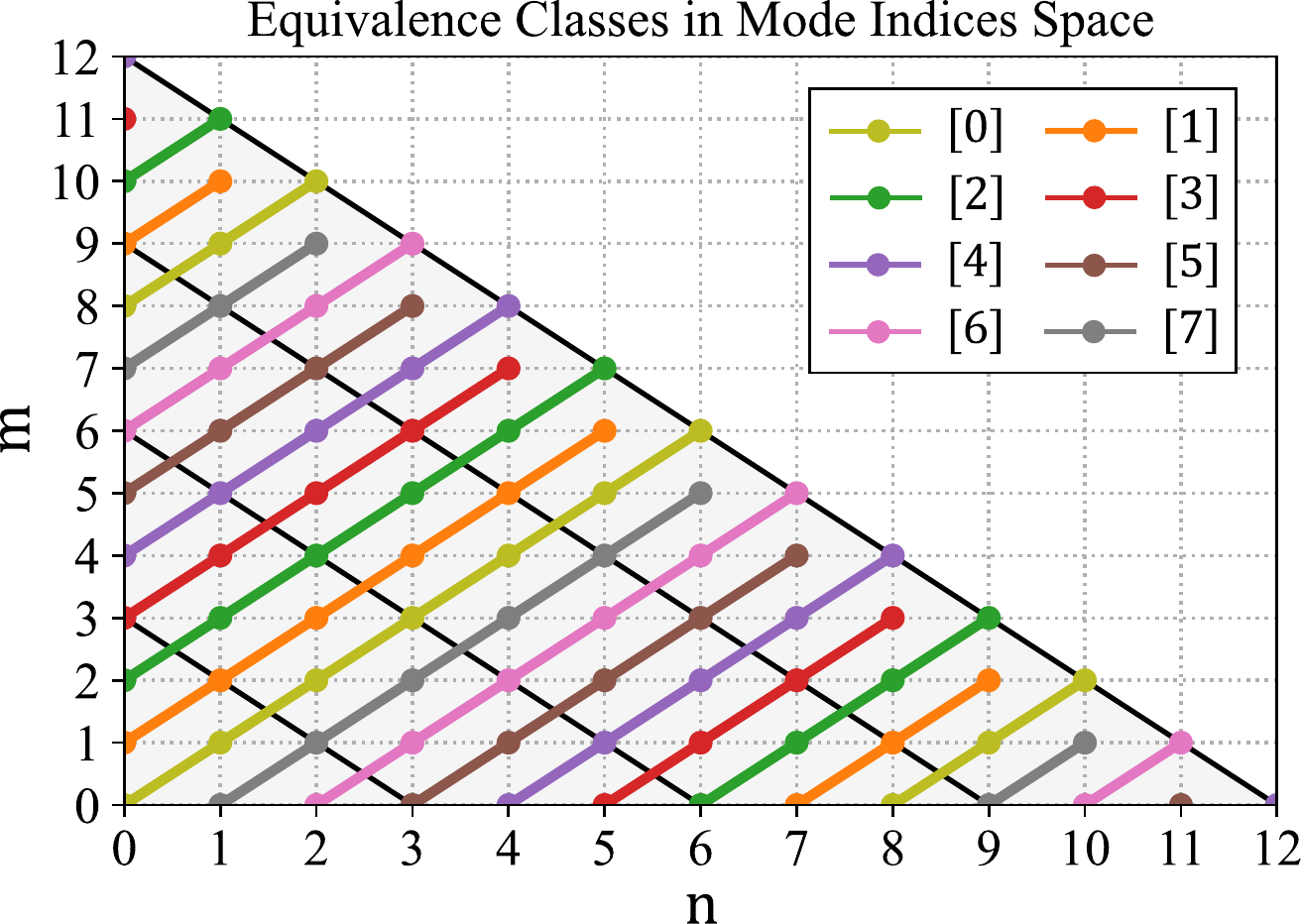}
	\caption{Visual representation of the equivalence classes. Black anti-diagonal lines represent modes of orders $N_{max}= 3, 6, 9,12 $. Light Grey background includes all modes up to the order $ N_{max}=12 $ for illustration purposes. Diagonal lines are visual guides for modes of the same equivalence class $ [j] $, represented by different colors.\label{fig:equi}}
\end{figure}

We can see in \cref{fig:equi} a visual representation of the equivalence classes in the mode indices space where every point point represents a different HG mode. The maximum mode order $ N_{max} $ sets an anti-diagonal line and the set of all included modes is represented as the shaded light gray area below the curve. The equivalence classes are separated by colors and \textit{diagonal lines are used as visual guides}. In summary, every point in a line of a given color is a mode of order $ m,n $ associated with an equivalence class $ [j] $.
\subsection{Generalization of Eigenvectors}

As demonstrated in the previous section, we can create superpositions of eigenvectors of the same eigenvalue, which themselves are eigenvectors.  We call these auxiliary eigenvectors. By inspection we can write them as

\begin{equation}\label{eq:eigenvec}
	\begin{split}
		\ket{\Psi}_j= \sum_{n=0}^{\lfloor \frac{N_{max}-j}{2} \rfloor }\sum_{\alpha=0}^{\lfloor\frac{N_{max}-j-2n}{8}\rfloor} c_{8\alpha+j+n}^n \ket{HG^n_{8\alpha+j+n}} \\	+\sum_{n=0}^{\lfloor\frac{N_{max}+j}{2}\rfloor}\sum_{\alpha=1}^{\lfloor\frac{N_{max}+j-2n}{8}\rfloor} c^{8\alpha+n-j}_n \ket{HG_n^{8\alpha+n-j}}
	\end{split}
\end{equation}
where $ \ket{\Psi}_j $ is the linear combination of eigenvectors associated with the eigenvalue $ \lambda_j $ and $ N $ is the modal order up to which modes will be considered. The first summation includes modes with $ m>n $ where the second one includes modes with $ n>m $. The $ \lfloor \circ \rfloor $ notation represents the floor function. The summation over HG modes can be done up to an arbitrary number of modes given that established relations are followed. Truncating the sum up until an arbitrary maximum mode order is done in order to have an estimation of mode sizes and propagation dynamics, according to \cite{Carter1980,Phillips1983}. From the Gouys phase know that the maximum mode order considered will dictate how "fast" (in propagation) the intensity profile will vary and the lowest order included will determine how "fast" the field will converge to its far-field pattern. Therefore, one can manipulate the propagation dynamics by excluding/including modes of given orders.

\begin{table}[h!]
	\centering
	\begin{tabular}{|l|c|l|} 
		\hline
		[j] & Eigenvalue & Mode Superposition \\ [0.5ex] 
		\hline\hline
		[0] & 1 & $ \left|\text{HG}_3^3\right\rangle +\left|\text{HG}_2^2\right\rangle +\left|\text{HG}_1^1\right\rangle +\left|\text{HG}_0^0\right\rangle $ \\
		\hline
		[1] & $ \sqrt[4]{-1} $ & $ \left|\text{HG}_2^3\right\rangle +\left|\text{HG}_1^2\right\rangle +\left|\text{HG}_0^1\right\rangle $ \\ 
		\hline 
		[2] & i & $ \left|\text{HG}^0_6\right\rangle +\left|\text{HG}_2^4\right\rangle +\left|\text{HG}_1^3\right\rangle +\left|\text{HG}_0^2\right\rangle $ \\
		\hline 
		[3] & $ (-1)^{3/4} $ &  $ \left|\text{HG}^0_5\right\rangle +\left|\text{HG}_1^4\right\rangle +\left|\text{HG}_0^3\right\rangle $  \\
		\hline 
		[4] & $ -1 $ & $ \left|\text{HG}^1_5\right\rangle +\left|\text{HG}_1^5\right\rangle +\left|\text{HG}^0_4\right\rangle +\left|\text{HG}_0^4\right\rangle $\\
		\hline 
		[5] & $ -\sqrt[4]{-1} $ & $ \left|\text{HG}^1_4\right\rangle +\left|\text{HG}_0^5\right\rangle +\left|\text{HG}^0_3\right\rangle $ \\
		\hline 
		[6] & $ -i $ &  $ \left|\text{HG}^2_4\right\rangle +\left|\text{HG}_0^6\right\rangle +\left|\text{HG}^1_3\right\rangle +\left|\text{HG}^0_2\right\rangle $ \\
		\hline 
		[7] & $ -(-1)^{3/4} $ &  $ \left|\text{HG}^2_3\right\rangle +\left|\text{HG}^1_2\right\rangle +\left|\text{HG}^0_1\right\rangle $ \\
		\hline
	\end{tabular}
	\caption{Table depicting a superposition of modes grouped by eigenvalue up to a order of $ N_{max}=6 $ with all coefficients set to $ 1 $.}
	\label{table:modes_n6}
\end{table}

Given a single equivalence class $ [j] $, if more than one mode is included, there is an infinite number of possible superpositions to be considered, since all $ c^n_m $ coefficients are continuous complex variables. Of course, the more HG modes are included in this summation, the more different the intensity patterns look from each other.  For simplicity, we will use all coefficients $ c^n_m=1 $ since the resilience character of this modes is independent of an initial phase and relative non-zero weights. Since the auxiliary eigenvectors are composed of orthogonal modes and each mode can only be associated with only one auxiliary eigenvector, they are also orthogonal among themselves, not mattering the choice of relative weights and/or phases. The \cref{table:modes_n6} explicitly summarizes the superpositions considered for examples in this work, with corresponding equivalence classes and eigenvalues up until $ N_{max}=6 $. The normalization factor was omitted.

\section{Experimental Setup}
\begin{figure*}
	\includegraphics[width=1\linewidth]{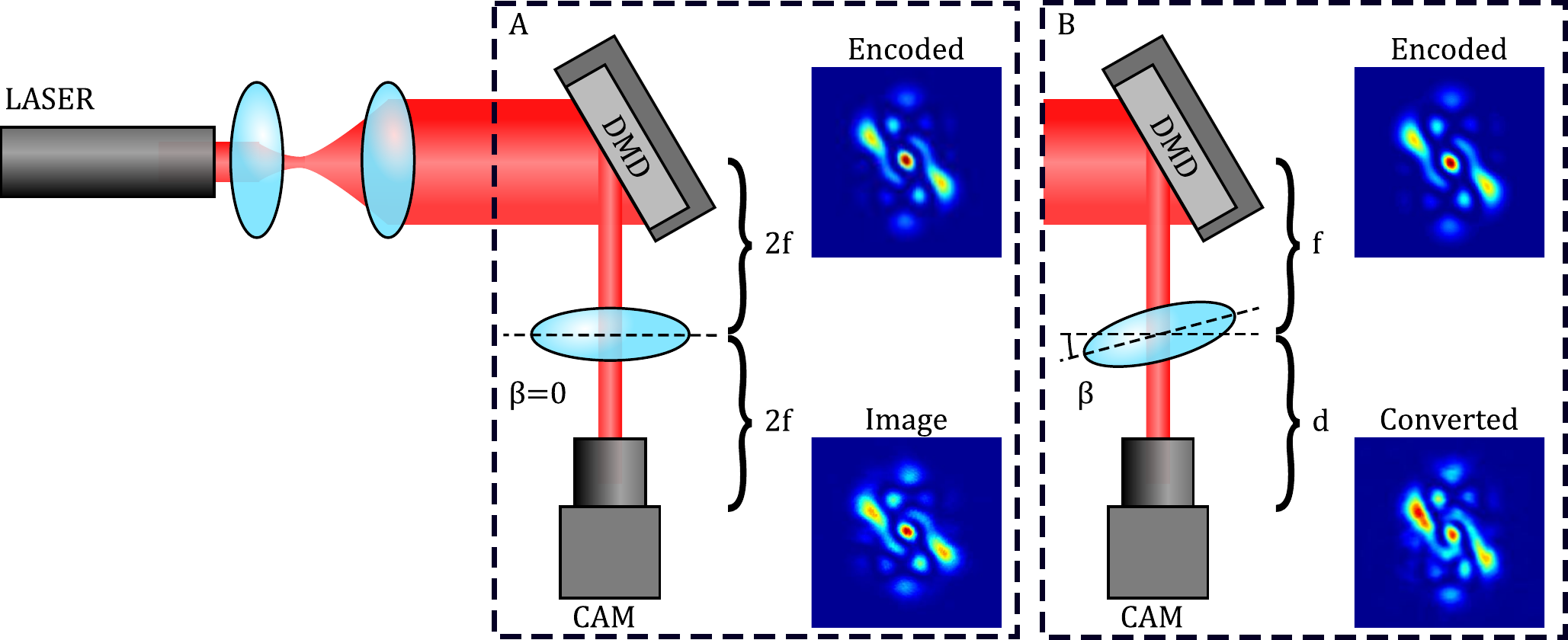}\caption{The experimental setup consists of a mode being generated on the DMD and subsequently being transformed by the tilted lens in a camera. A second arm with another lens is also set.\label{fig:exp}}
\end{figure*}

The experiment consists of a beam generation stage followed by a lens and a camera. The beam generation starts with the expansion and subsequent collimation of the beam, which impinges on a digital micromirror-device. In this device we encode binary holograms \cite{ren2015tailoring,scholes2019structured} which generates the desired field.  In order to confirm that we are generating the intended superposition of modes, we first build a $ 4f $ imaging system using a single lens, as depicted in \cref{fig:exp} situation A. This arrangement images the auxiliary eigenvectors right after being shaped by the DMD right into the CCD camera, where the pattern is confirmed to be the one encoded.

\vspace{0.5cm}

Afterwards, we know that the operation described by \cref{eq:operator} only works in a specific plane in propagation after the lens, so it is necessary to make slight alterations and move from situation A to B, as depicted in \cref{fig:exp}.
In a distance $ f $ of the DMD screen we put a lens of focal length $ f $. After this, we placed the camera in a distance $ d \approx f $ and tilt the lens.

\vspace{0.5cm}

The process to find the specific plane goes as follows: a LG mode is encoded on the DMD and confirmed to be observed. In a distance $ f $ after the DMD we place a lens and at a distance $ f $ after the lens the camera is placed. This constitutes a $ 2f $ system in which the far-field distribution is obtained. In the case of an LG, both near-field and far-field are equivalent. Then, the lens is rotated around the $ z $ axis by a $ \beta $ angle until the image observed on the camera is close to that of a HG mode oriented at $\pm 45 ^o $. This process requires some slight compensation of the propagation direction, as the tilting of the lens can redirect the beam. We subsequently move the camera closer to the lens until an exact image of a HG mode is observed. The measured distance $ d = 25  $ cm was measured between the lens and the camera when used a lens of $ f=30 $ cm with a measured inclination of $ \beta=19^o $.

With this we confirm that the observed plane is indeed the proper conversion plane. We then proceed by changing the encoded modes on the DMD to those of \cref{eq:eigenvec}. We demonstrate this by setting $ N_{max}=6 $ and generating the auxiliary eigenvectors according to \cref{table:modes_n6}. This is illustrated in \cref{fig:exp} situation B.

\subsection{Propagation}

\begin{figure*}
	\centering
	\includegraphics[width=1\linewidth]{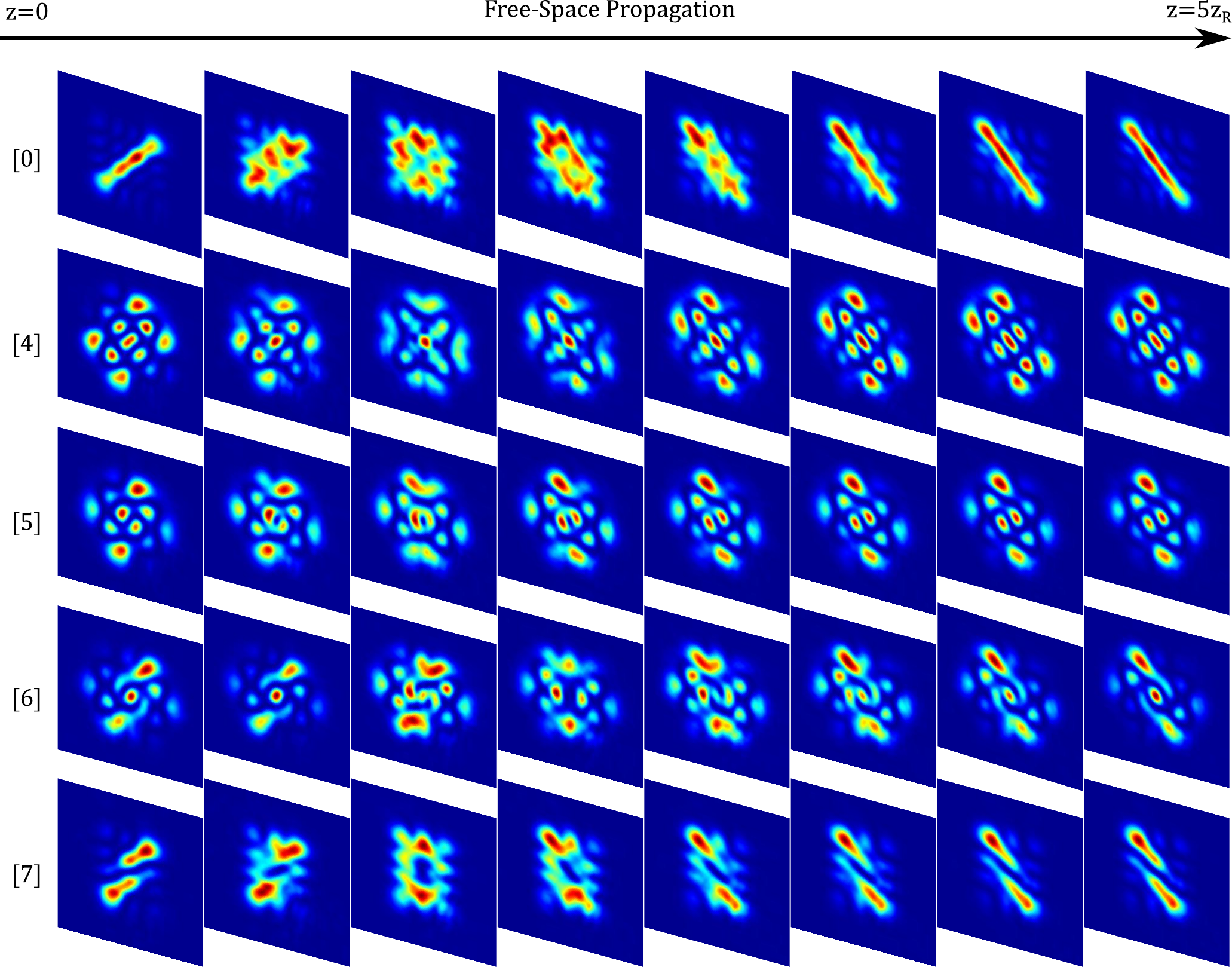}\caption{Propagation dynamics of NTEs of $ N_{max}=6 $.\label{fig:propag}}
\end{figure*}

While variant in free-space propagation, the dynamics of a superposition of modes is trivial: besides resizing, each mode picks up a phase according to their Gouy phases, which is proportional to the mode order. In the auxiliary eigenvectors intensity pattern this is translated into the interesting behaviour of the far-field being the mirror image of the near-field, a flip in the $ x $ direction. This is illustrated with experimental data in \cref{fig:propag}. It is also interesting to point out that, since the propagation phase is manifested as intramodal phases, the field in any point in propagation is still an auxiliary eigenvector. The operator of the tilted lens $ \mathcal{M}_t $ describes three processes: propagation to the lens, application of a astigmatic phase given by the tilting of the lens and the propagation to the conversion plane, as illustrated in \cref{fig:concept}. When considering single modes or superpositions of modes of the same order, the first propagation stage is negligible, as it only reescales the modes. In the case of a superposition of modes of different orders, the field that reaches the lens is not the original encoded near-field since the propagation phase each constituting mode picks up will be different. Therefore, when dealing with auxiliary eigenvectors, the initial propagation has to be taken into account. 

\vspace{0.5cm}

We dealt with this is two different ways: the first one is that, since we are dealing with paraxial modes, any phase propagation is just a relative phase between modes of different orders. We can compensate this by adjusting the encoded hologram so that we change the initial field in order to obtain a desired field after the tilted lens. 
The second way is to use a digital propagation technique \cite{Singh2021} whereby encoding a lens function in the phase of the encoded mode we shift the fourier plane. Both ways showed identical results in the conversion plane.

\vspace{0.5cm}

This claim is supported by the fact that in \cite{Vaity2013} the astigmatic mode conversion happens when the phase is applied \textit{to the field that reaches the lens}. For a field consisting of a single mode, the near-field and the one that reaches the lens are essentially the same, but this not true for superpositions of modes that propagate a non-negligible amount compared to the diffraction length.
Both methods were shown to compensate for the initial propagation phase. While the second one maintains the encoded intensity profile both in image of the near-field and conversion plane, the distance it would be able to compensate for is dependent on the resolution of the spatial light modulator, since the applied phase may vary very rapidly in the radial direction. The first method does not have this problem, as no extra phase is applied but instead the near-field is different from the field in the conversion plane. 

\vspace{0.5cm}

To further support our claims, we adapt our setup to that of a digitally applied astigmatic phase, by first setting $ \beta=0 $ (untilting the lens) and applying a phase proportional to the Zernike polynomial $ Z_2^2 $ associated with horizontal astigmatism. In this way, only the evolution under astigmatism is considered. The lens of focal distance $ f $ in a distance $ f $ after the DMD then creates a system that propagates the beam to its far-field at a distance $ f $, where the camera is placed. This situation is analogous to the field reaching the tilted lens without any propagation effects and then being propagated to the far field. 

\section{Results}

\begin{figure*}
	\centering
	\includegraphics[width=0.5\linewidth]{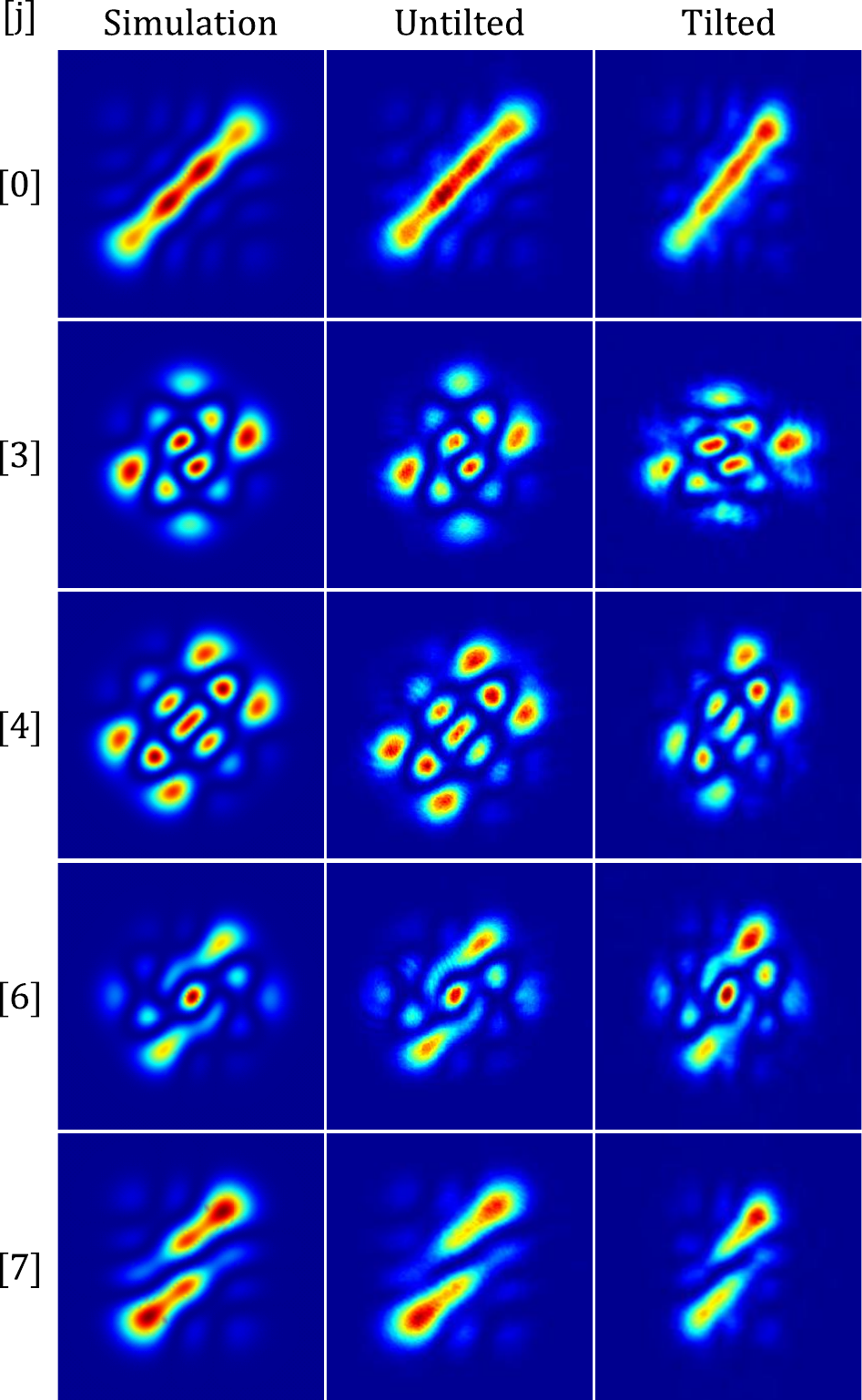}\caption{Experimental results for the modes convertion of modes grouped by $ j $ value up to the order $ N_{max}=6 $. First column shows theoretical simulations of the intensity profile. Second column shows the untilted image and third image shows the mode unaffected by the convertion process.  \label{fig:result1}}
\end{figure*}


In \cref{fig:result1} we can see results for a few modes of $ N_{max}=6 $. In the first column we show theoretical simulations of the encoded modes. In the second column (Untilted) are the near-field images obtained with the imaging setup described in \cref{fig:exp} situation A. The third column (Tilted) shows the modes after the tilted lens in the conversion plane, regarding \cref{fig:exp} situation B. It is possible to see remarkable agreement of simulations with experimental data: the originally encoded auxiliary eigenvectors are seen unaltered at the conversion plane of the tilted lens.

\vspace{0.5cm}

Notably, there is a lack of trivial symmetry. Modes with $ [j] \neq [0],[4] $ are not symmetric in either horizontal, vertical or $ \pm 45^\circ $ directions, while for $ [0],[4] $ there is inversion symmetry in $ \pm 45^\circ $. This can be explained by the fact that all constituting modes have different symmetries, which ultimately end up breaking one another. For the case of $ [0] $ one might observe that, since it follows that $ m-n=0 $, all modes will be symmetric in both $ x,y $ directions. This auxiliary eigenvector however can also include $ \ket{HG_0^8} $ and $ \ket{HG^0_8} $ which breaks the $ x,y $ symmetry but creates one at $ 45^\circ $. The same can be said for $ [4] $ which for every $ \ket{HG_{m'}^{n'}} $ also contains $ \ket{HG_{n'}^{m'}} $ and these two are the same, except for a $ 90^\circ $ rotation.  One might also notice that they have most of their intensities distributed along the $ 45^\circ $ line. This, however, is not related to astigmatic resilience, as we also know from \cite{ONeil2000} that a HG mode at $ 45^\circ $ is very similar but is not an eigenmode and gets converted into an LG mode.

\begin{figure*}
	\includegraphics[width=\linewidth]{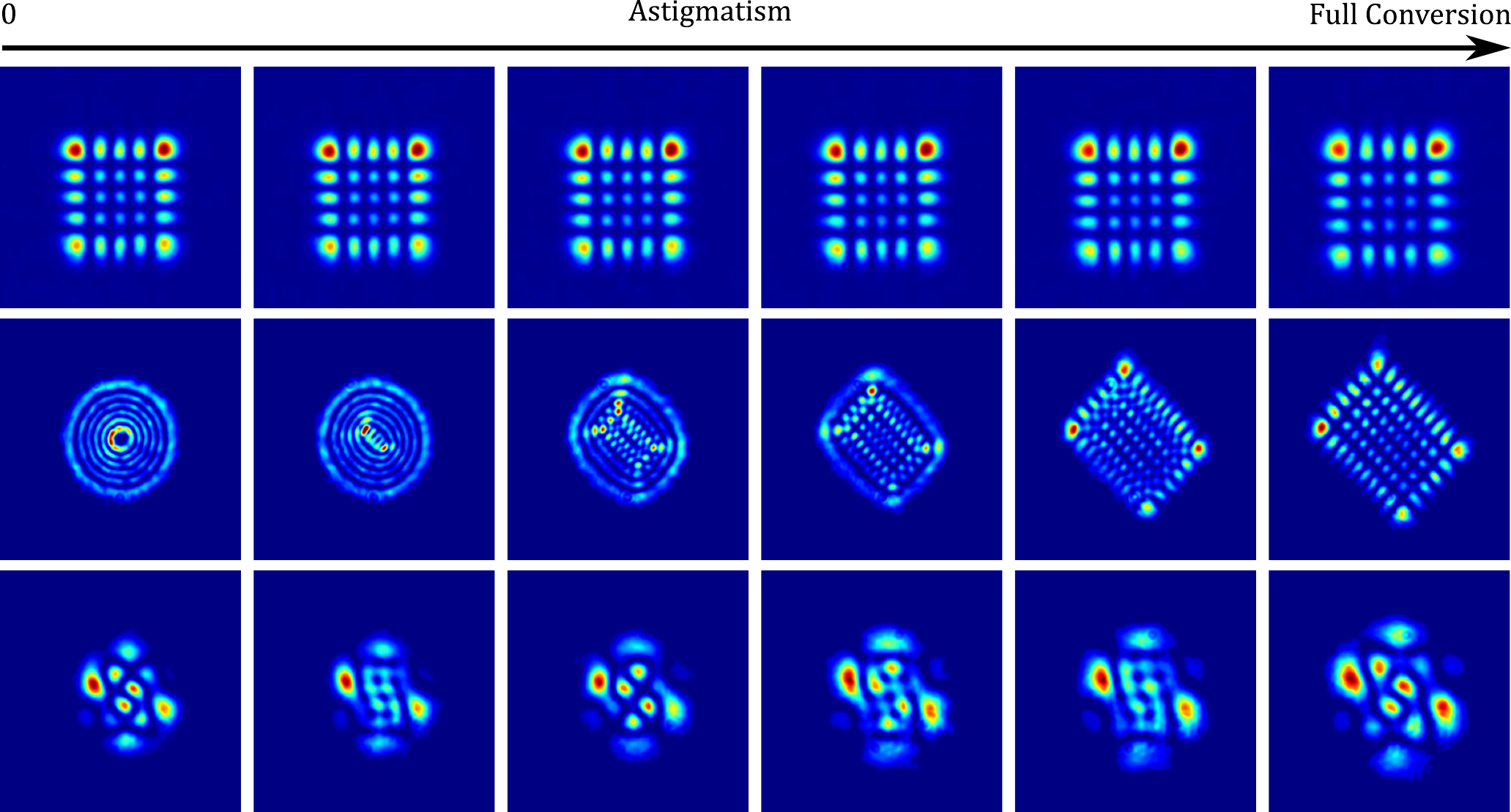}\caption{Experimental images of the beams when astigmatism is introduced in various degrees using Zernike polynomials. The first row shows the convertion of a high order LG mode and the second row shows the convertion of an auxiliary eigenvector.\label{fig:astig}}
\end{figure*}

\vspace{0.5cm}

Upon moving to the scheme where we can carefully increase the astigmatism of the system, we see the results in \cref{fig:astig}. There is no effect in a HG mode with $ m,n=4,4 $. Then we apply the astigmatism to a LG mode of $ l,p=5,-4 $. We can see that this mode is converted into a HG oriented at $ -45^o $ confirming the sign of the topological charge, as well as the indices $ m,n $ confirming the $ p,l $ indices according to \cite{Beijersbergen1993}. We then encode an auxiliary eigenvectors of $ N_{max}=6$ and $j=3 $ and show that, despite seeing changes when astigmatism is applied, the original profile is recovered for a level of astigmatism correspondent to full conversion.

\section{Conclusion}

Here we found the full set of eigenvectors of a highly aberrated channel, including the superpositions of eigenvectors, which can be grouped according to equivalence classes by a congruence relation to create new auxiliary eigenvectors. By truncating up to a maximum mode order we ensured that it is possible to control the physical dimensions and propagation dynamics. In this work we set the coefficients of all modes to be $ 1 $ for visualization, but this is not a requirement. These modes are not stable under propagation, but all show the behavior of the far-field being the flipped image of the near-field in the $ x $ axis. Interestingly, propagation for these modes is only an intra-modal phase given by the Gouy phase, which in turn does not change the validity of the eigenmode superposition still being an eigenmode. This implies that every intensity form seen in \cref{fig:propag} is also an auxiliary eigenmode which propagate independently of the astigmatism. However, we also showed that the level of astigmatism is an important condition to the full recovery of the original profile. 

\vspace{0.5cm}

We studied the astigmatic mode converter, which has a well defined operator with a given periodicity. However, this approach can be extended to any astigmatic optical system, given that it is unitary. By probing a system with a basis and generating a matrix in the modal space, it is not only possible to obtain eigenmodes of the system, but any periodicity of the eigenvalues would be detected if probed with a sufficient number of modes. Those can be grouped to form intricate patterns that are unaltered by this system and even exhibit properties beyond the original eigemodes while retaining the resilience property, such as the $ x $ flip behavior and symmetry manipulation.

\vspace{0.5cm}

In this system, the equivalence classes can have beams of many different orders. Of course, the size and type of equivalence classes depends heavily on the optical system, but this separation can be very broad and mutually non-exclusive with recent applications based on Gouy phase effects, such as \cite{Silva2021,Zhong2021}.


\medskip
\textbf{Acknowledgements} \par 
Wagner Tavares Buono acknowledges support from the CAPES institution for access to scientific publications, Dr. Braian Pinheiro da Silva for fruitful discussions and the creators of the Quantum Notation package for Wolfram Mathematica (Prof. José Luis Gómez-Muñoz and Dr. Francisco Delgado).

\medskip

%

\bibliographystyle{MSP}
\bibliography{eigen_tilt.bib}


\end{document}